# Higher-order Modes and Heating

*B.J. Holzer[1]*
CERN, Geneva, Switzerland

**Abstract**
This chapter gives a basic introduction to the problem of wake fields created in the beam-surrounding environment and the resulting heating effects of machine components. The concepts are introduced and scaling rules derived that are exemplified by several observations from operation of the LHC and other machines.

## 1 Introduction

In the treatment of transverse as well as longitudinal beam dynamics, the movement of the particles is described under the influence of the external fields, in general ignoring – or neglecting – the interaction of the particles with the fields created directly or indirectly by the beam itself.

The situation visualized in Fig. 1 is therefore highly idealized: it shows the result of a particle trajectory (green) following the usual rules of focusing and defocusing fields, within the beam envelope defined by the β-function (red) but without any influence from self or image currents [1]. And at least in the case of intense beams, these additional aspects have to be taken into account.

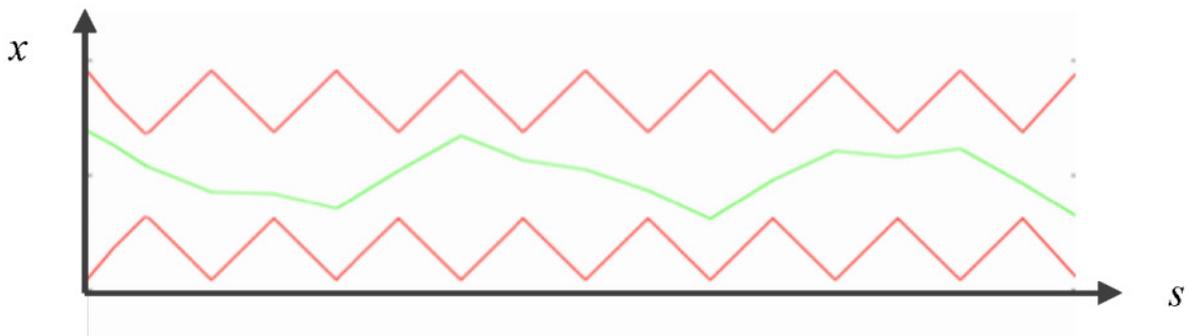

**Fig. 1:** A single-particle trajectory in a storage ring

Both the dynamics of individual particles as well as the collective dynamics of the complete particle distribution within a bunch and between different bunches depend greatly on the interaction between the electromagnetic field of the particles and their environment; that is, in general, the properties and geometry of the vacuum chamber. The treatment is usually based on the concept of impedance and therefore a hopefully useful reminder follows.

While in an ordinary electrical DC circuit, the potential and current are related by Ohm's law,

$$U = RI \, ,$$

the relation between an AC voltage with a frequency $\omega$ and phase $\phi_U$,

[1] bernhard.holzer@cern.ch

$$U_{ac} = U_0 \cos(\omega t - \phi_U), \tag{1}$$

and the corresponding current

$$I_{ac} = I_0 \cos(\omega t - \phi_I), \tag{2}$$

usually written more elegantly in complex notation, as

$$U_C = U_0 e^{(i\omega t - \phi_U)}, \; U_C = I_0 e^{(i\omega t - \phi_I)}, \tag{3}$$

is defined by the impedance $Z$:

$$Z = \frac{U_C}{I_C} = \frac{U_0}{I_0} e^{-(\phi_U - \phi_I)}. \tag{4}$$

It is convenient to use the same concept for the relation of the particle beam (representing just another AC current) and the voltage induced via the interaction of its fields with the beam-surrounding material; that is, the vacuum chamber.

Before doing so, a second reminder is added here. A charged particle in its rest frame is surrounded by an electric field that propagates isotropically in all directions. Being accelerated to larger and larger velocities, this field, observed in the laboratory frame, is Lorentz contracted and the field lines spread out longitudinally only within an angle of $\pm 1/\gamma$. While for some low-energy heavy ion rings the relativistic $\gamma$-parameter can still be quite low, we will restrict the description here to ultra-relativistic electron or proton rings. For example, in the case of the LHC running at 7 TeV beam energy, we obtain $1/\gamma = 1.3 \times 10^{-4}$.

Qualitatively, the situation obtained is shown in Fig. 2.

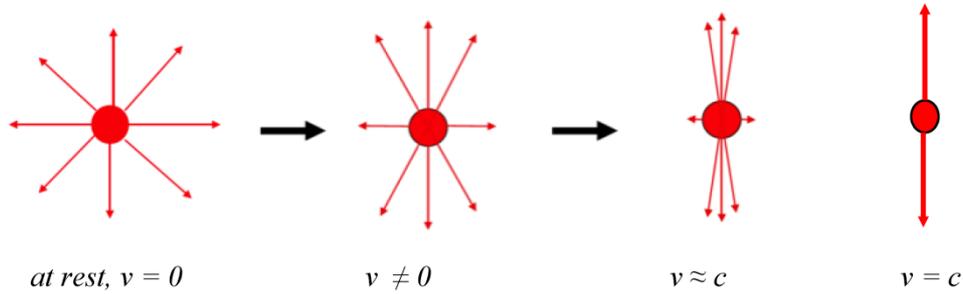

*at rest, $v = 0$*      *$v \neq 0$*      *$v \approx c$*      *$v = c$*

**Fig. 2:** The field lines of a particle at rest and moving at higher and higher velocities. (Courtesy of D. Brandt, in [2].)

Due to these transversely extending field lines, an image current is created that is floating inside the vacuum chamber walls, travelling with the bunch along the accelerator structure. In a uniform vacuum tube with perfectly conducting walls, the image current is floating without losses and no net forces are generated that would act back on the bunch. In summary, if an ultra-relativistic beam is travelling in a perfect orbit, in a perfectly conducting and perfectly smooth vacuum chamber, no heating effects are obtained (and there are no collective instabilities).

Unfortunately, these three conditions are not realistic. In particular, the vacuum chambers used in storage rings are designed following a number of different criteria, and they are far from being superconducting or smooth (even if, nowadays, '*smooth-fullness*' is increasingly part of a successful design).

Due to this non-ideal situation, the image charges are retarded, the field lines sketched in Fig. 2 are distorted, and the corresponding 'wake' fields act back on the beam (Fig. 3).

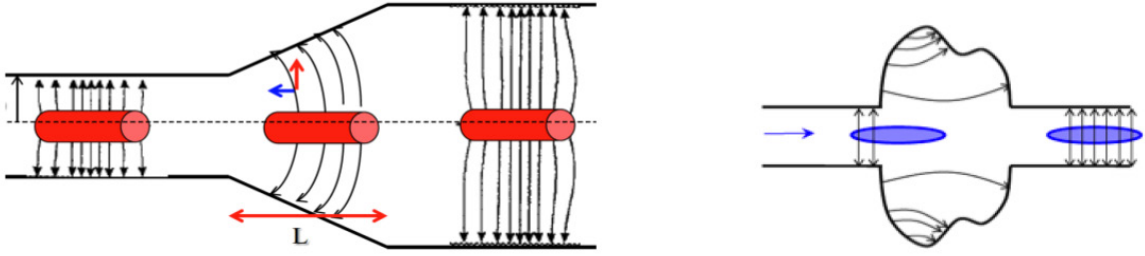

**Fig. 3:** The field lines of a particle bunch in a non-uniform vacuum chamber geometry for a sudden change of the chamber geometry and a cavity-like object.

Both effects – the resistivity of the chamber material and the field retardation – will lead to longitudinal components of the electrical field and as a consequence they will lead to energy loss of the beam and at the same time to heating effects in the vacuum system.

### 1.1 Wake fields and impedances

The interaction of the charged particle beam with its environment can be treated either in the time domain or in the frequency domain, and the two descriptions are completely equivalent.

In the time domain, the explicit fields are calculated that act back on the beam. In the frequency domain, the vacuum chamber components are treated as (frequency-dependent) impedances that describe the relation between the stored beam current and the corresponding voltage. A strong coupling between beam and environment is expected if the frequency spectrum of the bunched beam and the impedance of the surrounding chamber have significant components at the same frequencies. In accordance with the original concept of impedance (Eq. (4)), we set

$$V(\omega) = -Z(\omega)I(\omega), \tag{5}$$

where the minus sign indicates the decelerating effect, leading to energy loss of the particles. The impedance is, in general, a complex number and depends on the shape and material of each chamber component.

Further differentiation is useful in this context: for a particular component of the vacuum chamber, we can obtain wake fields due to sudden changes in the chamber cross-section – these typically have 'RF quality factors' of $Q \approx 1$ – or due to cavity-like objects (as on the right-hand side of Fig. 3) with $Q \gg 1$. As in the case of classical oscillators, the quality factor (describing the damping term of the system) and the bandwidth are related, and so we talk about narrow bandwidths for $Q \gg 1$ and broadband systems for $Q \approx 1$, where the quality factor $Q$ of the system is defined as the ratio between the stored energy in the cavity and the energy loss per period [3]:

$$Q = 2\pi \times \frac{\text{stored energy}}{\text{energy loss per period}}. \tag{6}$$

As an example, the resonance curves for three significantly different $Q$-factors, $Q = 1$, 10, and 100, in storage ring cavities are shown in Fig. 4. The extreme is obtained in superconducting cavities where, for example, for the LHC resonators, values as high as $Q = 8 \times 10^4$ are obtained, leading to an extremely narrow resonance curve.

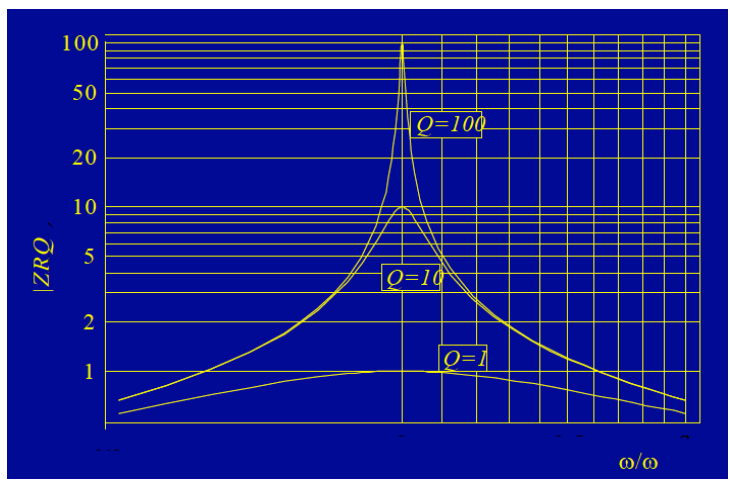

**Fig. 4:** Resonance curves for different quality factors, $Q$. (Courtesy of E. Jensen, in [3].)

The effect on the beam depends strongly on this $Q$-factor and thus on the details of the chamber geometry.

Cavity-like objects, with their high quality factors, have a narrow bandwidth and couple to the beam in only a small frequency band. However, the fields created by a bunch inside these objects will stay long enough to react on subsequent bunches or even – after one turn – back on the same bunch.

Low quality factors describe objects such as sudden changes of the vacuum chamber size; they have a broad frequency bandwidth but in general will not act on subsequent bunches. However, due to their frequency spectrum they will easily couple back on the bunch that created them.

An extreme example of the first case is given in Fig. 5: it shows the detector beam pipe of the HERMES experiment [4] that had been installed in the HERA collider. To achieve the highest luminosity and the best detector resolution, the beam pipe radius at the IP was reduced to about 1 cm, and careful tapering to the accelerator standard vacuum chamber was needed to limit the energy deposit due to the wake fields to within tolerable limits.

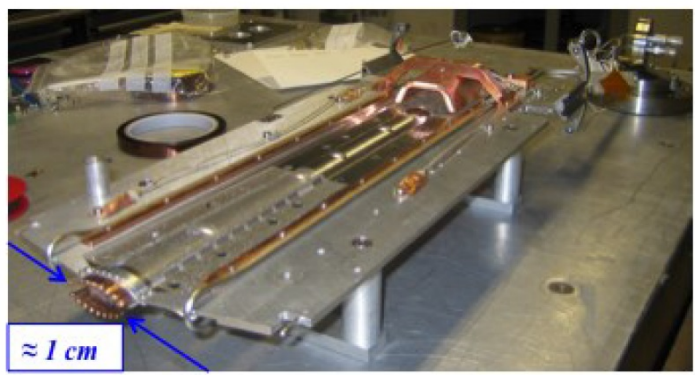

**Fig. 5:** The detector beam pipe of the HERMES experiment, representing a low $Q$-value object

An example of the second extreme in the case of the LHC collider is shown in Fig. 6: for the detection of particles created during the collision process and to close the detector acceptance as much as possible, additional detector components are installed (so-called 'forward spectrometers') that are moved on to the beam during stable luminosity operation. Due to the nature of the problem, they perfectly fulfil the conditions of a cavity-like object.

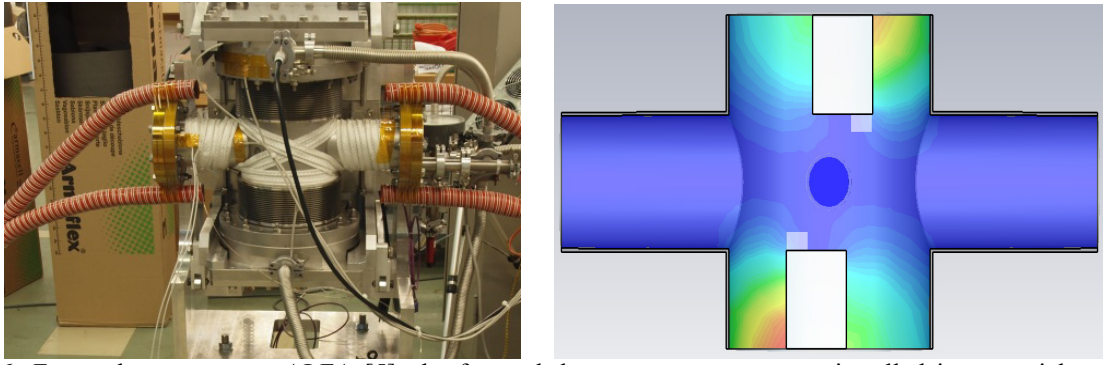

**Fig. 6:** Forward spectrometer ALFA [5]: the forward detector components are installed in a special vacuum device and are a good representation of a high *Q*-factor 'cavity-like' object for the beam.

## 1.2 Longitudinal wake fields

Following the qualitative picture of the wake fields, longitudinal and transverse fields can be generated if the bunch is passing a sudden change in the vacuum chamber geometry. While the transverse fields will deflect the beam and might lead to instabilities, the longitudinal wake fields will lead to energy loss of the particles and create heating of the vacuum components.

In order to obtain some kind of quantitative picture, we observe the situation of a test particle at the end of the bunch at position $\tilde{z}$, which will be influenced by the wake fields of the particles ahead of it. Each charge at position $z > \tilde{z}$, ahead of our test particle, will create wake fields that influence our test particle, as shown schematically in Fig. 7.

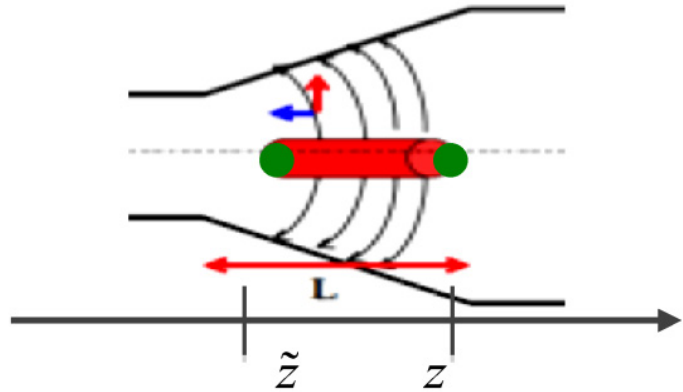

**Fig. 7:** A particle at the head of the bunch (position *z*) creates wake fields that act back on a test particle at the bunch tail (position $\tilde{z}$).

The longitudinal wake function is defined by the integral of the longitudinal field $E_\parallel$ and normalized to the charge *q* of the leading particle. Following the notation in Ref. [6], we write

$$W_\parallel(z-\tilde{z}) = \frac{1}{q} \int_L E_\parallel(s, t - \Delta z / \beta c) \mathrm{d}s, \qquad (7)$$

where, for causality reasons, only the cases $z - \tilde{z} > 0$ are relevant and *L* refers to the complete interaction length, which can be the length of the vacuum component or, in the extreme case, the complete circumference of the machine. The integral in Eq. (7) represents the longitudinal potential that is acting on the test particle and, accordingly, the units of the wake function are [W] = V/C. Integrating over all particles ahead of our test particle and multiplying by its charge *e* will result in the total energy loss of our test particle; in other words, we get the total induced voltage or the wake potential:

$$V_{\text{hom}}(\tilde{z}) = -e \int_{\tilde{z}}^{\infty} \lambda(z)\, W_{\|}(z - \tilde{z})\, \mathrm{d}z, \tag{8}$$

where $\lambda$ describes the line density of the leading particle charges and the negative sign indicates the decelerating character of the fields. The total energy loss of the bunch is then obtained by integrating over all slices $\mathrm{d}\tilde{z}$:

$$\Delta U_{\text{hom}}(\tilde{z}) = -\int_{-\infty}^{\infty} e\lambda(\tilde{z})d\tilde{z} \int_{\tilde{z}}^{\infty} \lambda(z)\, W_{\|}(z - \tilde{z})\, \mathrm{d}z. \tag{9}$$

In short, if we know the bunch charge and the geometry of the vacuum chamber, we can calculate (numerically in most cases) the longitudinal fields and thus the wake function, Eq. (7), and obtain the resulting energy loss via Eq. (9).

Replacing the charge distribution in Eq. (8) by an instantaneous current at position $\tilde{z}$

$$I(\tilde{z},t) = \hat{I}_0 e^{i(k\tilde{z} - \omega t)} \tag{10}$$

we obtain (referring, for simplicity, to one single mode $k$)

$$V_{\text{hom}}(\tilde{z},t) = -\frac{1}{\beta c} \int_{\tilde{z}}^{\infty} I(\tilde{z}, t + \frac{z - \tilde{z}}{\beta c})\, W_{\|}(z - \tilde{z})\, \mathrm{d}z. \tag{11}$$

Applying a Fourier transformation of the current, we finally get a relation between the potential and the beam current:

$$V_{\text{hom}}(t,\omega) = -I(t,\omega)\frac{1}{\beta c} \int_{-\infty}^{\infty} e^{-\frac{i\omega \Delta z}{\beta c}} W_{\|}(z - \tilde{z})\, d\Delta z \tag{12}$$

In full equivalence to the situation of an AC circuit (Eq. (4)), we can now define the longitudinal impedance in the frequency domain as the parameter that relates current and voltage in our system,

$$Z_{\|}(\omega) = \frac{1}{\beta c} \int_{-\infty}^{\infty} e^{-i\omega \Delta z / \beta c} W_{\|}(\Delta z)\, \mathrm{d}\Delta z, \tag{13}$$

and as before it has the unit *ohm* ($\Omega$). The longitudinal coupling impedance relates the beam current to the induced voltage that is created by the wake field and that acts back on the beam:

$$V_{\text{hom}}(t,\omega) = -I(t,\omega) Z_{\|}(\omega). \tag{14}$$

To quote from Ref. [6], where these relations are nicely described, "The impedance of the environment of our beam is the Fourier transform of the wake fields left behind by the particle charges in this environment."

There is still work to do, as the key point and source of the problems are the longitudinal field lines that have to be calculated, and in general this only can be done numerically.

As an example, we present in Fig. 8 a famous historical plot that shows these field lines for a resistive wall wake field generated by a point charge $q$ [7].

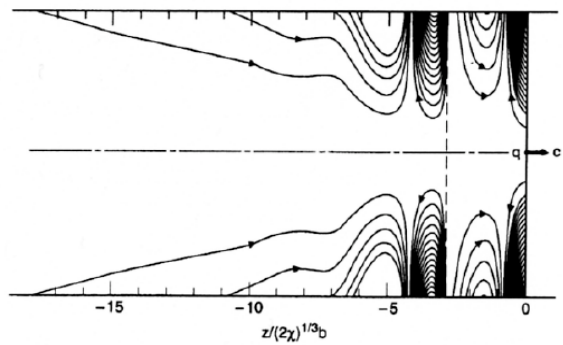

**Fig. 8:** The field lines for a resistive wall wake field generated by a point charge. (Courtesy of K. Bane.)

## 2    Observations on the beam

While in the end the total energy loss of the bunch is determined by the wake fields in the whole storage ring and thus the impedance of the complete system is of major importance for beam stability, large impedances of single accelerator components also have to be avoided. They might lead to local heating effects and can put severe limits on the operational performance.

During the commissioning phase of the LHC and the run 1 operation with an increasing number of bunches and single bunch intensities, several 'hot spots' in the true sense of the word were found.

An example of the effect of a sharp change in the aperture of the beam-surrounding vacuum system is shown in Fig. 9. It shows the temperature increase due to wake fields measured at a LHC beam collimator. By definition, these devices have to represent the locations with the tightest aperture in the storage ring, as they are supposed to protect any other machine components from unwanted beam losses. To reduce the problems of longitudinal wake fields, the collimator jaws are usually tapered to provide a smooth transition between the standard vacuum chamber and the collimator. Nevertheless, in the case of non-perfect tapering or other imperfections, wakes can be created at these places and will lead to strong localized energy deposits. The plot shows, during an LHC run lasting for seven days, a sharp temperature rise during beam injection and ramp, which falls slowly during collision runs due to the decreasing beam intensity. It has to be emphasized that the peak values in the plot refer to 50°C and were considered as the maximum tolerable limit for the system.

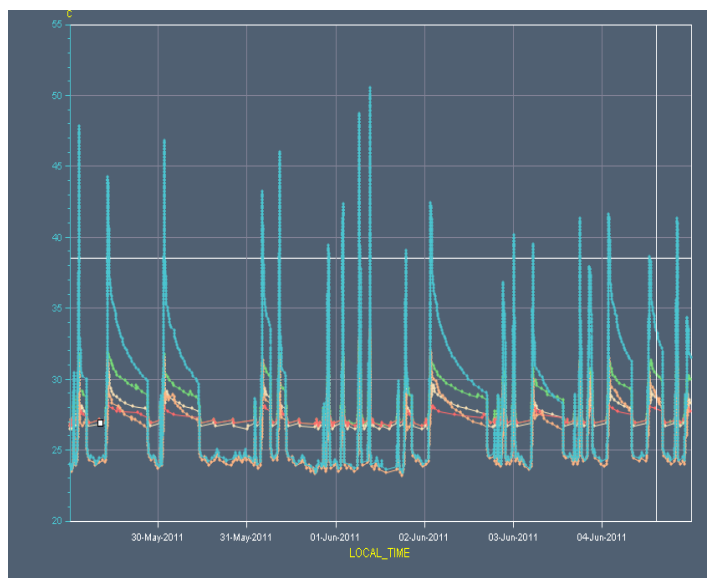

**Fig. 9:** The temperature increase measured at a LHC collimator (TCT) during injection, ramp, and stored beam

A quite similar effect has been observed at the 'TDI' collimator, which is positioned in the injection region of the LHC [8] and acts as beam stopper in case of injection failures. Due to local energy losses and the resulting heating effects, the RF shielding system that should provide a smooth beam environment has been deformed to such a degree that replacement of the system with an improved design became unavoidable (see Fig. 10).

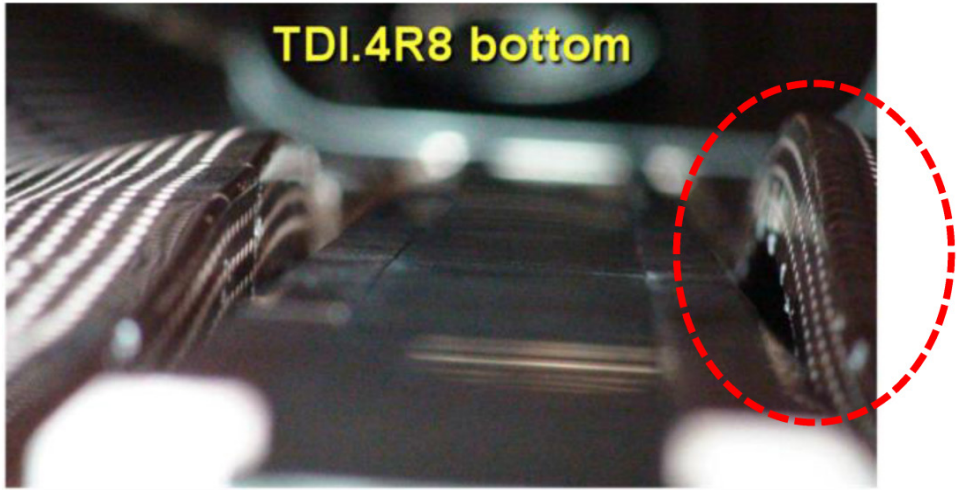

**Fig. 10:** The shielding of the LHC beam stopper, damaged due to wake field effects

In a real accelerator, the contribution of a large number of single elements to the overall wake field effects will add up. A parameter that is widely used to describe this integrated effect and that is a property of the complete vacuum system is the so-called 'energy loss factor'. It is defined as

$$k_{\text{hom}} = \frac{\Delta U_{\text{hom}}}{e^2 N_b^2} \quad (15)$$

where $\Delta U_{\text{hom}}$ describes the total energy loss of the bunch due to the wake field effect and $N_b$ is the number of particles in the bunch. The loss factor is clearly related to the wake function (Eqs. (7) and (9)) and we can write

$$k_{\text{hom}} = \frac{1}{N_b^2} \int_{-\infty}^{\infty} \lambda(\tilde{z}) d\tilde{z} \int_{\tilde{z}}^{\infty} \lambda(z) W_{\parallel}(z - \tilde{z}) dz. \quad (16)$$

As a general parameter of the system, the loss factor is of considerable importance as it determines – as we have seen above – the possible heating effects. The complete power loss of the beam is then obtained by

$$P_{\text{hom}} = k_{\text{hom}} \frac{I_0^2}{f_0 n_b} \quad (17)$$

with $n_b$ indicating the number of stored bunches, $f_0$ the revolution frequency, and $I_0 = n_b e N_b f_0$ the overall stored beam current. Just like losses in a conventional ohmic resistive system, the losses due to wake effects rise quadratically with the stored beam current.

Two remarks are needed in this context.

- The luminosity optimization of a storage ring collider depends on the beam current $I_0$ and bunch number $n_b$ (see Fig 11), and to optimize for the highest luminosities, the stored beam current should be distributed over a small number of bunches, $n_b$. This, however, leads in the end to performance limits due to energy deposit via increased wake effects, and to overcome this problem a compromise has to be found between optimum luminosity and tolerable heating effects.

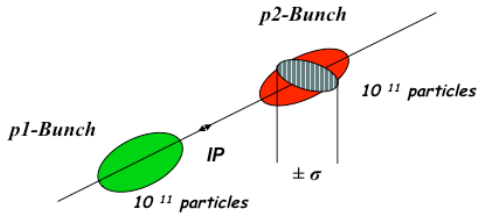

$$L = \frac{1}{4\pi e^2 \cdot f_0 \cdot n_b} \frac{I_0^2}{\sigma_x \cdot \sigma_y} \tag{18}$$

**Fig. 11:** Colliding bunches and luminosity in a collider

– The loss factor depends strongly on the bunch length. Figure 12 shows a historical measurement performed at the SPEAR collider.

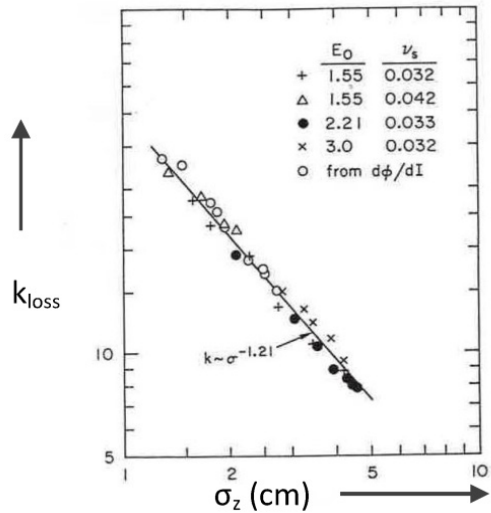

**Fig. 12:** The loss factor as a function of the bunch length, measured at SPEAR [9]

This bunch length effect has also been observed in the LHC, and a dedicated lengthening of the bunches has been applied, using RF noise, to limit the energy deposit in the critical systems. Figure 13 shows data taken within two subsequent LHC fills of equal beam parameters (such as the energy, intensity, and number of stored bunches) but for different bunch lengths. The curves show the beam current (green), the energy (yellow), the bunch length (red), and the collimator temperature (blue). In the first fill, the bunch is kept sufficiently long and the resulting temperature increase is quite limited. In the second fill, the bunch lengthening has been switched off and the bunch shortening that occurs naturally during acceleration leads to a sharp increase in the temperature of the collimator jaws, and finally to a beam dump triggered by the temperature interlock.

To avoid the limitations due to excessively short bunches in LHC operation, the natural shortening obtained during the acceleration, which depends on the energy as well as on the cavity voltage $\sigma_z \propto E^{1/4}$, $\sigma_z \propto V^{1/2}$, is counteracted by dedicated RF noise. Figure 14 shows the development of the bunch length during a LHC standard acceleration process. While at the start of the ramp a sharp decrease of the length is observed, careful 'shaking' of the bunches leads to a very reproducible bunch length of 1.2 ns at flat top.

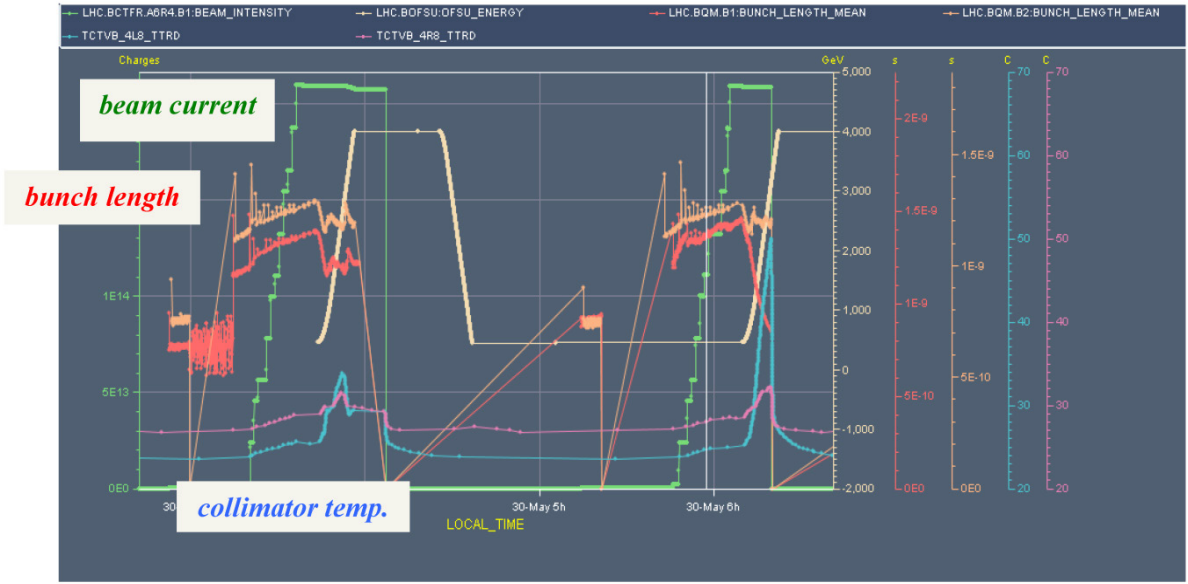

**Fig. 13:** The bunch length effect on the loss factor observed in the LHC

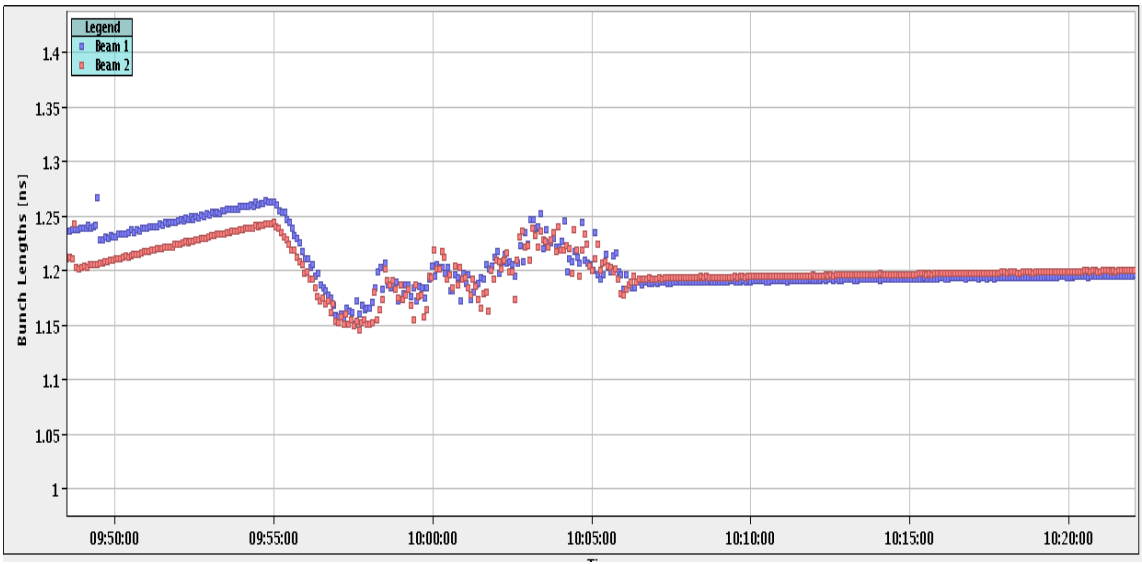

**Fig. 14:** The bunch length at the LHC during the acceleration process: the shortening effect is compensated in a dedicated way to achieve 1.2 ns at the end of the ramp.

## 3 Phase shift due to impedance effects

The energy loss caused by the wake field effect, which is described by the loss factor

$$k_{\text{hom}} = \frac{\Delta U_{\text{hom}}}{e^2 N_b^2}, \qquad (19)$$

will have to be compensated by the RF system of the storage ring and, due to the phase focusing principle (see Ref. [10]), leads to a shift in the synchronous phase of the beam. As a reminder, we repeat the qualitative argumentation here (Fig. 15).

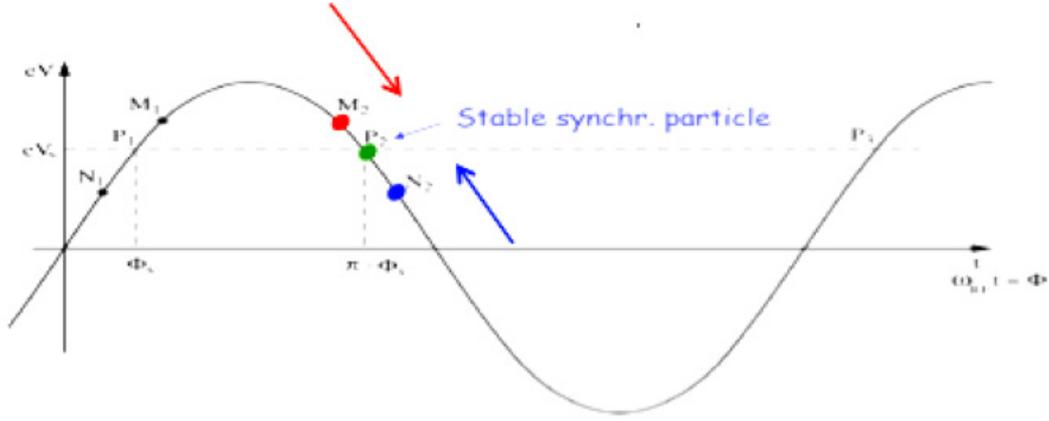

**Fig. 15:** The principle of phase focusing and the definition of the synchronous particle

The phase of the synchronous particle, marked by the green spot in the figure, is defined by the equilibrium between the average energy loss (due to synchrotron radiation, wake field effects etc.) and the energy gain from the RF system.

In the absence of impedance, the power gain of the bunch is given by

$$\Delta P_0 = e\hat{V}_{\rm rf} \cdot f_0 N_b \cdot \sin\phi_{\rm synch\_0}, \tag{20}$$

where $\hat{V}_{\rm rf}$ is the RF peak voltage and $\phi_{\rm synch\_0}$ describes the ideal synchronous phase. The impedance effects lead to an additional power requirement and, as a consequence, to a new phase $\phi_{\rm synch\_1}$. Accordingly, we obtain

$$\begin{aligned}\Delta P &= e\hat{V}_{\rm rf} \cdot f_0 N_b \cdot (\sin\phi_{\rm synch\_1} - \sin\phi_{\rm synch\_0}) \\ &\approx e\hat{V}_{\rm rf} \cdot f_0 N_b \cdot \Delta\phi_{1,2} \cdot \cos\phi_{\rm synch\_1}\end{aligned}. \tag{21}$$

Measurement of the shift in the synchronous phase can thus be used directly, to determine the loss factor of the machine.

## 4  Impedances in an accelerator

For completeness, we will briefly summarize some analytical expressions for the impedances in typical accelerator structures. A more detailed derivation can be found in Ref. [11].

### 4.1  Cavity-like structures

While the resonators used in an accelerator clearly present a dominant contribution to the impedance budget, it is hard to avoid them and the only optimization we can carry out is to reduce higher-order modes that will be the source of unwanted effects.

For the modelling of 'cavity-like objects', however, we can refer to the description that is typically used to characterize the effect of resonators in an accelerator, namely an LRC circuit (Fig. 16) with capacity $C$, inductivity $L$, and shunt impedance $R$, which are related to each other via

$$L = \frac{R_{\rm s}}{Q\omega} \quad \text{or} \quad \sqrt{\frac{L}{C}} = \frac{R_{\rm s}}{Q}. \tag{22}$$

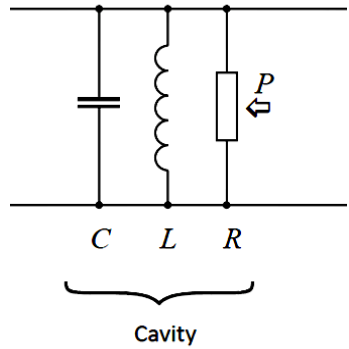

**Fig. 16:** An LCR circuit to describe the impedance of a resonator in a storage ring [2]

The impedance obtained is the sum of these three contributions:

$$\frac{1}{Z_\parallel} = \frac{1}{R_s} + \frac{1}{\omega L} - i\omega C, \qquad (23)$$

which we can write as

$$Z_\parallel = \frac{R_s}{1 + iQ\left(\frac{\omega_r}{\omega} - \frac{\omega}{\omega_r}\right)}. \qquad (24)$$

In order to avoid unwanted cavity-like structures to the maximum possible extent, a smooth vacuum system is needed, and whenever this is not possible so-called 'RF fingers' are installed to provide an RF-tight transition between vacuum components of different apertures. As an example, Fig. 17 shows such a device as used in the LHC storage ring.

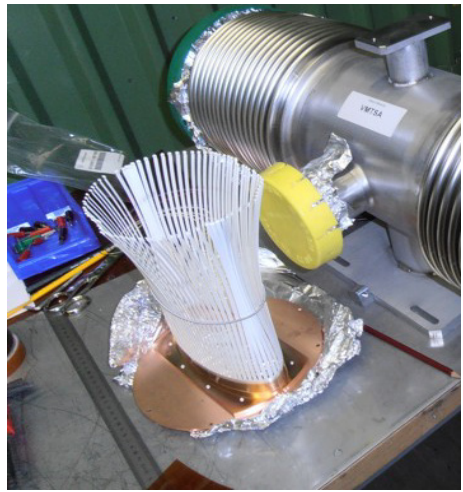

**Fig. 17:** The LHC interconnect device: elastic metal strips are used to provide an 'RF-tight' transition between different vacuum components and to avoid longitudinal wake fields.

Careful design of the vacuum chamber in terms of the geometry and the choice of materials is a must if high beam currents are desired (in other words, high machine performance). Standard techniques for the field calculations and construction have been developed and the largest part of the impedance budget can usually be understood and optimized. However, special devices exist in an accelerator where even today optimization is not free from problems. As an example, Fig. 18 shows the injection kicker for the LHC.

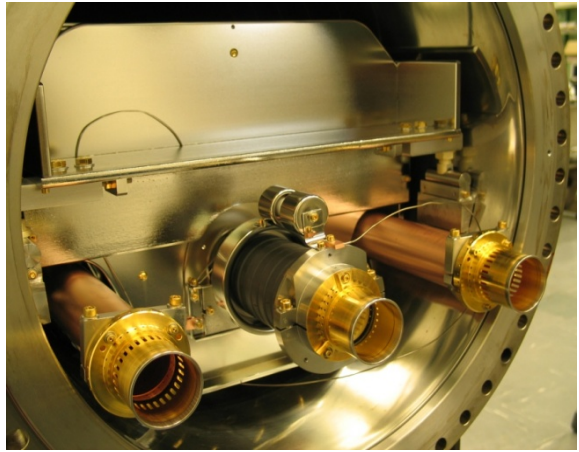

**Fig. 18:** The LHC injection kicker, with the ceramic beam tube in the centre

Due to the rapidly rising magnetic field that is needed to inject the new bunch trains from the transfer line into the storage ring, the vacuum chamber of the kicker magnets cannot be built out of conducting materials such as copper, steel, and so on. Eddy currents would be created in the surface of the kicker chamber and would perfectly shield the inside from the external magnetic field. As a consequence, the usual material chosen is ceramic, which is not really optimal, given the impedance and local heating effects. Ultimately, the kicker chamber and the surrounding vessel will form a beautiful 'cavity-like' object. A compromise therefore has to be found and the solution is to install a number of metallic wires, the purpose of which is just to deliver sufficient RF tightness and still allow the penetration of the kicker field to deflect the beam. As can be expected, an optimum has to be found – in the theoretical calculations, but also ultimately on the beam.

Figure 19 shows a comparison between the calculated and measured impedances for two different shielding configurations; one based on 15 wires, the other using 19. The reduction in impedance using a better shielding concept is as remarkable as the agreement between theory and measurement [12].

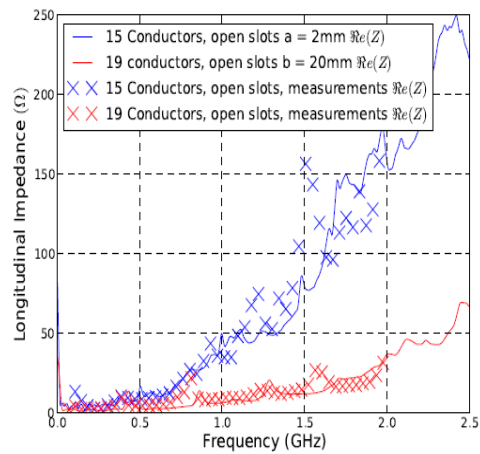

**Fig. 19:** Measured and calculated impedances for the LHC injection kicker

## 4.2 Resistive wall impedance

As explained above, the stored beam in an accelerator induces image currents in the vacuum chamber walls, which travel with the beam. For non-ideal conductivity, however, these image currents create resistive effects (i.e. losses in the chamber material) and lead to decelerating forces acting on the particles. These forces are proportional to the beam current itself, and integrating around the storage ring we get the expression [6]

$$\frac{Z_{\parallel}(\omega_n)}{n} = 1 + i\frac{\bar{R}}{nr_w \sigma_c \delta_{skin}} \qquad (25)$$

with $r_w$ describing the vacuum chamber radius, $\delta_{skin}$ and $\sigma_c$ the skin depth and conductivity of the material, and $n = \omega_n/\omega_0$ the frequency in units of the revolution frequency.

For a given geometry of the accelerator and the vacuum system, the only parameter that is left for optimization is the conductivity of the chamber material. In the LHC, therefore, a thin layer of copper has been used to cover the beam screen in order to reduce the resistive wall effects (see Fig. 20).

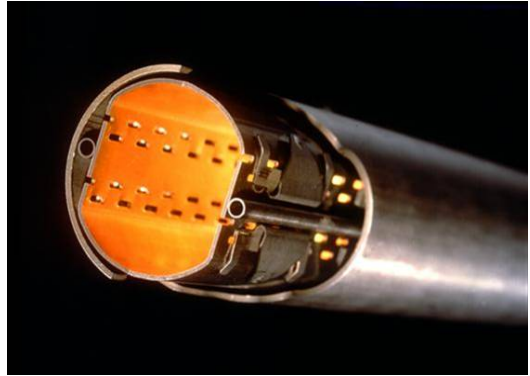

**Fig. 20:** The LHC beam screen: a thin layer of copper has been used to cover the surface, to reduce the resistive wall wake field effects.